\documentclass[12pt,hyper]{JHEP3}
\usepackage{graphics,psfrag}
\usepackage{amsmath,amsthm,amssymb,epsfig,euscript,array,cite,cancel}
\usepackage{cases,empheq}
\theoremstyle{definition}

\theoremstyle{remark}

\newcounter{multieqs}

\newcommand{\be}{\begin{equation}}
\newcommand{\ee}{\end{equation}}
\newcommand{\eq}[1]{(\ref{#1})}
\newcommand{\bit}{\begin{itemize}}  \newcommand{\eit}{\end{itemize}}

\newcommand{\bm}[1]{\mbox{\boldmath $#1$}}
\newcommand{\rf}[1]{(\ref{#1})}

\def\bd{\begin{document}}
\def\ed{\end{document}}
\def\nn{\nonumber}
\def\bea{\begin{eqnarray}}
\def\eea{\end{eqnarray}}
\let\bm=\bibitem

\def\la{\langle}
\def\ra{\rangle}

\def\npb#1#2#3{Nucl. Phys. {\bf{B#1}} #3 (#2)}
\def\plb#1#2#3{Phys. Lett. {\bf{#1B}} #3 (#2)}
\def\prl#1#2#3{Phys. Rev. Lett. {\bf{#1}} #3 (#2)}
\def\prd#1#2#3{Phys. Rev. {D \bf{#1}} #3 (#2)}
\def\cmp#1#2#3{Comm. Math. Phys. {\bf{#1}} #3 (#2)}
\def\cqg#1#2#3{Class. Quantum Grav. {\bf{#1}} #3 (#2)}
\def\nppsa#1#2#3{Nucl. Phys. B (Proc. Suppl.) {\bf{#1A}}#3 (#2)}
\def\ap#1#2#3{Ann. of Phys. {\bf{#1}} #3 (#2)}
\def\ijmp#1#2#3{Int. J. Mod. Phys. {\bf{A#1}} #3 (#2)}
\def\rmp#1#2#3{Rev. Mod. Phys. {\bf{#1}} #3 (#2)}
\def\mpla#1#2#3{Mod. Phys. Lett. {\bf A#1} #3 (#2)}
\def\jhep#1#2#3{J. High Energy Phys. {\bf #1} #3 (#2)}
\def\atmp#1#2#3{Adv. Theor. Math. Phys. {\bf #1} #3 (#2)}

\def\N{{\cal N}}
\def\sst{\scriptscriptstyle}
\def\thetabar{\bar\theta}
\def\Tr{{\rm Tr}}
\def\one{\mbox{1 \kern-.59em {\rm l}}}

\def\a{\alpha}      \def\da{{\dot\alpha}}  \def\dA{{\dot A}}
\def\b{\beta}       \def\db{{\dot\beta}}  
\def\g{\gamma}  \def\G{\Gamma}  \def\dc{{\dot\gamma}}  
\def\d{\delta}  \def\D{\Delta}  \def\ddt{\dot\delta}  
\def\e{\epsilon}        \def\ve{\varepsilon}  
\def\f{\phi}    \def\F{\Phi}    \def\vvf{\f}  
\def\h{\eta}  
\def\k{\kappa}  
\def\l{\lambda} \def\L{\Lambda}  
\def\m{\mu} \def\n{\nu}  
\def\o{\omega}  
\def\p{\pi} \def\P{\Pi}  
\def\r{\rho}  
\def\s{\sigma}  \def\S{\Sigma}  
\def\t{\tau}  
\def\th{\theta} \def\Th{\Theta} \def\vth{\vartheta}  
\def\X{\Xeta}  
\def\z{\zeta}  

\def\na{\nabla}  

\def\cA{{\cal A}} \def\cB{{\cal B}} \def\cC{{\cal C}}  
\def\cD{{\cal D}} \def\cE{{\cal E}} \def\cF{{\cal F}}  
\def\cG{{\cal G}} \def\cH{{\cal H}} \def\cI{{\cal I}}  
\def\cJ{{\cal J}} \def\cK{{\cal K}} \def\cL{{\cal L}}  
\def\cM{{\cal M}} \def\cN{{\cal N}} \def\cO{{\cal O}}  
\def\cP{{\cal P}} \def\cQ{{\cal Q}} \def\cR{{\cal R}}  
\def\cS{{\cal S}} \def\cT{{\cal T}} \def\cU{{\cal U}}  
\def\cV{{\cal V}} \def\cW{{\cal W}} \def\cX{{\cal X}}  
\def\cY{{\cal Y}} \def\cZ{{\cal Z}}

\def\ua{\underline{\alpha}}  
\def\uc{\underline{\phantom{\alpha}}\!\!\!\gamma}  
\def\um{\underline{\mu}}  
\def\ud{\underline\delta}  
\def\ue{\underline\epsilon}  
\def\una{\underline a}\def\unA{\underline A}  
\def\unb{\underline b}\def\unB{\underline B}  
\def\unc{\underline c}\def\unC{\underline C}  
\def\und{\underline d}\def\unD{\underline D}  
\def\une{\underline e}\def\unE{\underline E}  
\def\unf{\underline{\phantom{e}}\!\!\!\! f}\def\unF{\underline F}  
\def\unm{\underline m}\def\unM{\underline M}  
\def\unn{\underline n}\def\unN{\underline N}  
\def\unp{\underline{\phantom{a}}\!\!\! p}\def\unP{\underline P}  
\def\unq{\underline{\phantom{a}}\!\!\! q}  
\def\unQ{\underline{\phantom{A}}\!\!\!\! Q}  
\def\unH{\underline{H}}  
  
\def\As {{A \hspace{-6.4pt} \slash}\;}  
\def\bs {{b \hspace{-6.4pt} \slash}\;}  
\def\Ds {{D \hspace{-6.4pt} \slash}\;}
\def\Gts {{\Gt \hspace{-6.4pt} \slash}\;}
\def\ds {{\del \hspace{-6.4pt} \slash}\;}  
\def\ss {{\s \hspace{-6.4pt} \slash}\;}  
\def\ks {{ k \hspace{-6.4pt} \slash}\;}  
\def\ps {{p \hspace{-6.4pt} \slash}\;}   
\def\xs {{x \hspace{-6.4pt} \slash}\;}  
\def\pas {{{p_1} \hspace{-6.4pt} \slash}\;}  
\def\pbs {{{p_2} \hspace{-6.4pt} \slash}\;}   
\def\cFs {{{\cal F} \hspace{-6.4pt} \slash}\;}

\def\Ah{{\hat{A}}}  
\def\Dh{{\hat{D}}}
\def\Gh{{\hat{G}}}
\def\Fh{{\hat{F}}}
\def\Ih{{\hat{I}}} 
\def\Jh{{\hat{J}}} 
\def\Kh{{\hat{K}}}
\def\Lh{{\hat{L}}} 
\def\Ph{{\hat{P}}}
\def\Rh{{\hat{R}}}
\def\Vh{{\hat{V}}} 
\def\Xh{{\hat{X}}}
 
\def\ah{{\hat{\a}}}
\def\bh{{\hat{\b}}}
\def\gh{{\hat{\g}}}
\def\dh{{\hat{\d}}}
\def\hh{\hat{h}}
\def\uh{\hat{u}}  
\def\xh{\hat{x}}  
\def\yh{\hat{y}}  
\def\ph{\hat{p}}  
\def\xih{\hat{\xi}}  
\def\chih{\hat{\chi}}  
\def\Psih{\hat{\Psi}}

\def\psit{\tilde{\psi}}  
\def\Psit{\tilde{\Psi}}   
\def\Psibt{\tilde{\bar{Psi}}}  

\def\st{\tilde{\sigma}}  

\def\delt{\tilde{\delta}}
\def\Phit{\tilde{\Phi}}   
\def\Phitb{\overline{\tilde{Phi}}}  
\def\tht{\tilde{\th}}  
\def\lt{\tilde{\l}}
\def\chit{\tilde{\chi}}   
\def\phit{\tilde{\phi}} 

\def\At{\tilde{A}}
\def\Bt{\tilde{B}}
\def\Ct{\tilde{C}}
\def\Dt{\tilde{D}}
\def\Et{\tilde{E}}
\def\Ft{\tilde{F}}
\def\Gt{\tilde{G}}
\def\Ht{\tilde{H}}
\def\It{\tilde{I}}
\def\Jt{\tilde{J}}
\def\Qt{\tilde{Q}}  
\def\Rt{\tilde{R}}  
\def\Mt{\tilde{M }}  
\def\Nt{\tilde{N}}   
\def\St{\tilde{S}}
\def\Vt{\tilde{V}}
\def\Xt{\tilde{X}} 
\def\at{\tilde{a}}
\def\ct{\tilde{c}}
\def\dt{\tilde{d}}
\def\htt{\tilde{h}} 
\def\ft{\tilde{f}}
\def\gt{\tilde{g}}
\def\pt{\tilde{p}}  
\def\qt{\tilde{q}}  
\def\vt{\tilde{v}}  
\def\nt{\tilde{n}}  
\def\ut{\tilde{u}}  
\def\wt{\tilde{w}}  
\def\zt{\tilde{z}} 
\def\xt{\tilde{x}} 
\def\yt{\tilde{y}} 
\def\Psit{\tilde{\Psi}}
\def\vphit{\tilde{\varphi}}  

\def\eb{\bar{\epsilon}} 
\def\delb{\bar{\partial}}  
\def\thb{\bar{\theta}}
\def\mub{\bar{\mu}}
\def\lamb{\bar{\l}}
\def\psib{\bar{\psi}}
\def\sb{\bar{\sigma}}
\def\xib{\bar{\xi}}
\def\chib{\bar{\chi}}

\def\Psib{\bar{\Psi}}
\def\Phib{\bar{\Phi}}
\def\Lamb{\bar{\Lambda}}
\def\Sb{{\overline \Sigma}}
\def\cb{\bar{c}}
\def\hb{\bar{h}}
\def\qb{\bar{q}}
\def\wb{\bar{w}}
\def\ub{\bar{u}}
\def\zb{{\bar{z}}}
\def\Hb{\bar{H}}
\def\Qb{{\bar Q}}
\def\Omegab{\overline{\Omega}}
\def\ob{\overline{\omega}}

\def\Ab{{\overline A}} \def\Bb{{\overline B}} \def\Cb{{\overline C}}  
\def\Db{{\overline D}} \def\Eb{{\overline E}} \def\Fb{{\overline F}}  
\def\Gb{{\overline G}} 
\def\Ib{{\overline I}}  
\def\Jb{{\overline J}} \def\Kb{{\overline K}} \def\Lb{{\overline L}}  
\def\Mb{{\overline M}} \def\Nb{{\overline N}} \def\Ob{{\overline O}}  
\def\Pb{{\overline P}}  \def\Rb{{\overline R}}  
 \def\Tb{{\overline T}} \def\Ub{{\overline U}}  
\def\Vb{{\overline V}} \def\Wb{{\overline W}} \def\Xb{{\overline X}}  
\def\Yb{{\overline Y}} \def\Zb{{\overline Z}}  

\def\fb{{\overline f}}
\def\gb{{\overline g}}
\def\mb{{\overline m}}
\def\lb{{\overline l}}
\def\yb{{\overline y}}
  
\def\ldel{{\overleftarrow{\del}}}
\def\rdel{{\overrightarrow{\del}}}
\def\ldeldel{{\overleftarrow{\del^2}}}
\def\rdeldel{{\overrightarrow{\del^2}}}
\def\ldelb{{\overleftarrow{\bar{\del}}}}
\def\rdelb{{\overrightarrow{\bar{\del}}}}

\def\ba{{\bf a}} 
\def\bk{{\bf k}}  
\def\bl{{\bf l}}  
\def\bp{{\bf p}}  
\def\bq{{\bf q}}  
\def\br{{\bf r}}
\def\bt{{\bf t}}
\def\bu{{\bf u}}
\def\bv{{\bf v}}
\def\bx{{\bf x}}  
\def\by{{\bf y}}  
\def\bR{{\bf R}}  
\def\bV{{\bf V}}

\def\bone{{\bf 1}}  

\def\va{{\vec a}}
\def\vk{{\vec k}}
\def\vp{{\vec p}}
\def\vq{{\vec q}}
\def\vx{{\vec x}}
\def\vy{{\vec y}}
\def\vu{{\vec u}}
\def\vv{{\vec v}}

\def\vs{{\vec \sigma}}
\def\vtau{{\vec \tau}}

\newcommand{\ov}[1]{\overrightarrow{#1}}

\def\frA{\mathfrak{A}}
\def\frB{\mathfrak{B}}
\def\frC{\mathfrak{C}}
\def\frD{\mathfrak{D}}
\def\frE{\mathfrak{E}}
\def\frF{\mathfrak{F}}
\def\frG{\mathfrak{G}}
\def\frH{\mathfrak{H}}
\def\frM{\mathfrak{M}}
\def\frN{\mathfrak{N}}
\def\frR{\mathfrak{R}}
\def\frW{\mathfrak{W}}

\def\fra{\mathfrak{a}}
\def\frb{\mathfrak{b}}
\def\frf{\mathfrak{f}}
\def\frg{\mathfrak{g}}
\def\frh{\mathfrak{h}}
\def\frl{\mathfrak{l}}
\def\frs{\mathfrak{s}}
\def\fri{\mathfrak{i}}
\def\frj{\mathfrak{j}}

\def\ma{\mathfrak{a}}
\def\mg{\mathfrak{g}}
\def\mh{\mathfrak{h}}
\def\mR{\mathfrak{R}}
\def\mN{\mathfrak{N}}

\def\cHt{\tilde{\cH}}

\def\d{\delta}\def\D{\Delta}\def\ddt{\dot\delta}  
  
\def\pa{\partial} \def\del{\partial}  
\def\xx{\times}  
\def\uno{\mbox{1 \kern-.59em {\rm l}}}    
  
\def\trp{^{\top}}  
\def\inv{^{-1}}  
\def\dag{{^{\dagger}}}  
\def\pr{^{\prime}}  
  
\def\rar{\rightarrow}  
\def\lar{\leftarrow}  
\def\lrar{\leftrightarrow}  
  
\newcommand{\0}{\,\!}      
\def\one{1\!\!1\,\,}  
\def\im{\imath}  
\def\jm{\jmath}  
  
\newcommand{\tr}{\mbox{tr}}  
\newcommand{\slsh}[1]{/ \!\!\!\! #1}  
  
\def\vac{|0\rangle}  
\def\lvac{\langle 0|}  
  
\def\hlf{\frac{1}{2}}  
\def\ove#1{\frac{1}{#1}}  

\def\Box{\square}  
\def\CC {\mathbb{C}}
\def\FF {\mathbb{F}}
\def\RR{\mathbb{R}}
\def\NN{\mathbb{N}}  
\def\ZZ{\mathbb{Z}}  
\def\bb#1{{\bf #1}}  
\def\bcomment#1{}  
\def\bfhat#1{{\bf \hat{#1}}}  
\def\VEV#1{\left\langle #1\right\rangle}  

\newcommand{\ex}[1]{{\rm e}^{#1}} \def\ii{{\rm i}}  

\newcommand{\lrbrk}[1]{\left(#1\right)}
\newcommand{\sfrac}[2]{{\textstyle\frac{#1}{#2}}}
 
\def\stw{{\sqrt{2}}}

\def\rf {{\rm f}}
\def\ri {{\rm i}}
\def\rj {{\rm j}}
\def\rk {{\rm k}}
\def\rl {{\rm l}}
\def\rs {{\scriptscriptstyle \rm S}}
\def\rt {{\scriptscriptstyle \rm T}}

\def\rQ {{\scriptscriptstyle \rm \cQ}}
\def\rR {{\scriptscriptstyle \rm \cR}}

\def\cQb{{\cal \Qb}}
\def\cRb{{\cal \Rb}}
\def\cWb{{\cal \Wb}}

\def\fd {{\rm N}}
\def\afd {{\overline{\rm N}}}

\def \II {I\hspace{-.1em}I\hspace{.1em}}
\def \IIA {\mbox{\II A\hspace{.2em}}}
\def \IIB {\mbox{\II B\hspace{.2em}}}
\def \gs {g^s}
\def \ls {\lambda^s}

\def \I {{\cal I}}
\def \qs {q\hspace{-.53em}/\hspace{.15em}}
\def \ks {k\hspace{-.53em}/\hspace{.15em}}
\def \YM {{\mbox{\tiny YM}}}
\def \gym {g_{\YM}}

\def \Lc {\L_c}
\def\IR{\relax{\rm I\kern-.18em R}}
\def \id {{\bf 1}}

\def\cci{\ell}
\def\ccj{\ell'}

\def \thbb{\overline{\th\th}}
\newcommand \ol{\overline}
\def \lamb{\bar{\lambda}}
\def \vphi{\varphi}
\def \lambh{\hat{\bar{\lambda}}}
\def \lh{\hat{\lambda}}
\def \dd{\ddagger}

\def \Xd{\dot{X}}
\def \nd{\noindent \\}

\author{Chong-Sun Chu\\  
Centre for Particle Theory
and Department of Mathematical Sciences,\\ 
Durham University, Durham, DH1 3LE, UK \\
E-mail:  
\email{chong-sun.chu@durham.ac.uk} }

\title {A Theory of 
Non-Abelian Tensor Gauge Field with Non-Abelian Gauge Symmetry $G\times G$}

\abstract{
The Chern-Simon action of the ABJM theory is not gauge invariant 
in the presence of
a boundary. In the paper 
\href{http://arxiv.org/abs/arXiv:0909.2333}{\cite{CS2}}, this was 
shown to imply 
the existence of a Kac-Moody
current algebra
on the theory of multiple
self-dual strings. In this paper we 
conjecture that the Kac-Moody symmetry induces a $U(N)\times U(N)$ 
gauge symmetry in the theory of $N$ coincident M5-branes. As a start, 
we construct a $G\times G$ gauge symmetry algebra structure which
naturally includes the tensor
gauge transformation for a non-abelian 2-form tensor gauge field. 
The gauge covariant field strength is constructed.
This new $G\times G$ gauge symmetry algebra allows us to write down a theory of 
a non-abelian tensor gauge field 
in any dimensions. The $G \times G$ gauge bosons can be either
propagating, in which case the 2-form gauge fields would interact with
each other through the 1-form gauge field; or they can be auxiliary and
carry no local degrees of freedom, in which case the 2-form gauge
fields would be self-interacting nontrivially. 
We finally comment on the possible application to the system of 
multiple M5-branes.
We  note that the field content
of the  $G\times G$ non-abelian tensor gauge theory can be 
fitted nicely into
(1,0) supermultiplets; and we suggest a construction of 
the theory of multiple M5-branes with manifest (1,0) supersymmetry.

}
\preprint{DCPT-11/43}
\keywords{M-Theory, D-branes, M-branes, Gauge Symmetry}

\begin{document}

\section{Introduction}

The low energy theory of $N$  coincident M5-branes is given by
an interacting (2,0) superconformal  theory in 6 dimensions. 
So far very little is known about
this theory.  As a first step, one would like to
understand what kind of gauge symmetry
structure underlies the worldvolume theory of multiple M5-branes. 
This is the primary goal of this paper.

Great progress has been made in the last couple of years 
for the case of multiple M2-branes. 
First a new class of (2+1)-dimensional superconformal field theories
with maximal $\cN=8$ supersymmetry was constructed 
by Bagger and Lambert \cite{BL1,BL2,BL3}, 
and by Gustavsson \cite{Gut}. The construction makes use of a new
mathematical object called a Lie 3-algebra which is a generalization
of 
the Lie
algebra. However the
application to describe multiple M2-branes has been hindered by a 
major difficulty that so far there is only one example of a Lie
3-algebra that could produce a well defined unitary quantum theory. 
Another proposal due to Aharony, Bergman, Jafferis and Maldacena 
\cite{ABJM} 
proposed  a certain $\cN=6$ superconformal 
Chern-Simons-matter theory as  the low energy theory of multiple
M2-branes.
In this construction, an ordinary $U(N) \times U(N)$ Lie algebra is used 
and the rank $N$ is arbitrary.
It has been argued that the inclusion of a nonperturbative monopole
sector enhances the  supersymmetry to $\cN=8$ \cite{susy1,susy2,susy3,susy4}.  
   
Much less is known for the theory of multiple M5-branes. A possible approach to
this problem is to consider the M2-branes ending on the M5-brane(s) and to
make use of the recently obtained knowledge of multiple 
M2-branes to learn about the physics of M5-brane(s) from the boundary 
dynamics of the M2-branes. This approach has been
applied in \cite{CS1} and \cite{CS2}. In \cite{CS1}, the open BLG theory 
is considered and a novel kind of quantized geometry for M5-brane in a constant
$C$-field is predicted. In \cite{CS2}, a system of open $N$ M2-branes  
described by the 
open ABJM theory is considered. 
Due to the gauge non-invariance of the Chern-Simon
actions in 
the presence of a boundary, additional degrees of freedom 
must reside at the boundary of the M2-branes, which one could interpret
as the worldsheet of a system of multiple self-dual strings. 
These degrees of freedom is govern by a WZW  action on the group manifold
$\cG = U(N)\times U(N)$ and admits 
a Kac-Moody  $\cG_L\times \cG_R$ current algebra \cite{W2}. 
The existence of a Kac-Moody
current algebra is interesting and naturally one wonders what it
implies for the physics of the M5-branes, the spacetime of the
self-dual strings. 
This form the motivation and the starting point of the analysis
of this paper. 

In the literature, there has been various attempts in constructing a non-abelian
theory for the 2-form potential $B$. One class of attempts which also
involve the use of 1-form gauge fields is to use a mathematical 
structure called non-abelian gerbes \cite{gerbes}. Our construction is different
as some of the mathematical properties 
required in the non-abelian gerbes are not imposed in our construction. 
These properties 
(for example, the  vanishing of the fake curvature
as required in non-abelian gerbes in order to have a well defined 
parallel transport) 
are often well motivated mathematically, but 
their necessity are much less clear physically. 
As a result, the gauge transformations of the 2-form potential $B$ 
are different, for example.
Another class of attempt is to use a 
lattice definition of the tensor gauge connection. 
Interestingly, this line of proposal also automatically 
contains a $G \times G'$ gauge structure \cite{lattice}. 
For other recent works on the construction of the non-abelian 
(2,0) theory see \cite{lam1,dou,lam2,ho}.

In this paper we propose that the  Kac-Moody currents generate a $G\times G$
($G=U(N)$) gauge symmetry on the theory of $N$ coincident
M5-branes. We also propose to identify this gauge symmetry with the
non-abelian tensor gauge symmetry on multiple M5-branes. An immediate
question is how could a tensor gauge symmetry get generated from
Yang-Mills gauge symmetry? We find that if the $G\times G$ gauge
symmetry is not of the usual form but admits a kind of ``cross''
structure, then a tensor gauge transformation is automatically
included. 
This gauge symmetry structure allows us to write down immediately 
a theory of non-abelian  tensor gauge fields in any spacetime dimensions.
It is this $G\times G$ 
non-abelian tensor gauge symmetry algebra 
that we conjecture to be  
the symmetry of the 
low energy worldvolume theory 
of multiple M5-branes. 
  
Depending on the physical needs, the $G \times G$ gauge bosons can be 
constructed to be either propagating or non-propagating.
In the first possibility, the gauge bosons may obey, for example, 
a standard Yang-Mills term. In this case 
the 2-form gauge fields would interact with
each other through the 1-form gauge field. This is similar to the 
interaction of
fermions with gauge field in a minimally coupled theory. In the second
possibility, the gauge fields 
carry no local degrees of freedom and are determined entirely in terms of the
2-form potentials and other fields of the theory, and the 2-form gauge
fields would be self-interacting nontrivially. This second  possibility 
is particularly interesting
for the construction of the theory of multiple 
M5-branes as there is no room for
a propagating gauge field in the worldvolume supersymmetric multiplet of
M5-branes.

The plan and results of the paper are explained as follows.
In section 2.1 we argue that the Kac-Moody currents generate a $G\times G$
($G=U(N)$) gauge symmetry structure on the system of $N$ coincident M5-branes. 
In section 2.2, we introduce a set of $G\times G$ gauge bosons that is
characterized by a new set of  
gauge transformation laws that  are different from the standard
direct-product structure of gauge groups
\footnote{Since we will not use the standard direct-product gauge group 
structure and so there is no risk of confusion, we 
will use the same notation  $G\times G$ for our gauge symmetry structure.
}. 
In fact the algebra of the gauge transformations does not close by itself
and it is  necessary to include the tensor 
gauge transformation. Thus tensor gauge transformation is naturally and 
automatically included. 
A gauge covariant and tensor gauge invariant 
non-abelian 3-form field 
strength is constructed. 
We also discuss the possible physical natures of the gauge fields.
We also explain in what sense the full $G \times G$ gauge symmetry is essential 
in the construction of the non-abelian tensor gauge transformation.
In section 3, we discuss couplings to matter fields and 
show how to construct an invariant action in general dimensions. 
We also show how our non-abelian tensor gauge symmetry can be used to
construct a dual description of the 5-dimensional Yang-Mills gauge theory
of 1-form gauge potential $\cA_\m$ 
in terms of a non-abelian 2-form potential $B_{\m\n}$. 
The paper is concluded with some further discussions.
In particular, we briefly 
comment on  the application of our formalism of tensor
gauge symmetry to the construction of the self-dual
theory of multiple M5-branes.
We also note that the field content of the 
$G\times G$ non-abelian tensor gauge theory can be 
fitted nicely into (1,0) supermultiplets and  
we suggest that it may be more feasible to write down the non-abelian (2,0)
tensor theory of multiple M5-branes in terms of (1,0)
supermultiplets. Progress in these directions will be reported
elsewhere \cite{chua,chub}. 
Finally, we emphasis that a common feature in all these proposals 
\cite{dou,lam2,ho,chua,cg} is that
the proposed M5-branes theory is based on a gauge group $G=U(N)$.
This implies that the on-shell degrees of freedom scales like 
$N^2$ at large $N$. Whether and how an $N^3$ entropy scaling \cite{m5-S}
would arise is an interesting
question and deserve further investigations.

{\bf Note added}: During the preparation of this manuscript, the
  preprint \cite{sezgin} appeared 
which overlaps with some of the ideas of this
  paper. For example, both papers  
  draw on the similarity with the construction of ABJM and  suggests to
  construct the multiple M5 theory using the (1,0)
  supermultiplets. Also, both papers make use of the ordinary
  Lie algebra in describing the symmetry.  
However, the details of the constructions are different. 
For example, our
construction  is based on a special 
$G\times G$ ($G= U(N)$) gauge symmetry algebra that is 
not  in \cite{sezgin} and 
the rank $N$ is allowed to be arbitrary. 

\vskip 0.2cm 
{\bf Note added in v3}:
Recently, the authors of \cite{sezgin} (version 2) have checked that
by switching off the 3-form gauge potential in the tensor hierarchy
they proposed, the $G \times G$ symmetry proposed in this paper
provides a non-trivial solution to their construction. They also
obtained a set of (1,0) superconformal equations of motion from their
general construction and found that
these equations cannot be obtained from an action. 
Nevertheless it 
may still be possible to construct a supersymmetric action 
which is non-Lorentz
covariant or Lorentz invariant  
if one allows for new auxiliary field of PST type in the action. These
possibilities deserve further investigation \cite{chua,chub}.

\section{$G\times G$
Gauge Symmetry 
of Non-Abelian  Tensor Gauge Field}
 
\subsection{ $U(N) \times U(N)$ gauge symmetry on M5-branes}

Consider a system of  $N$ open M2-branes 
ending on $N$ coincident M5-branes. This can be modelled with 
the $U(N)\times U(N) \equiv \cG$ ABJM theory with boundary, together
with a certain
coupling to the non-abelian $B$-field living on the M5-branes. The
explicit form of this coupling is unknown, but the details are not
necessary for our argument.
It was shown in \cite{CS2} that the gauge non-invariance of the boundary
Chern-Simon couplings in the ABJM theory implies the existence of a $U(N)
\times U(N)$ WZW action for the multiple self-dual string theory.
In turn, this induces a  $\cG_L\times \cG_R$
Kac-Moody symmetry on the worldsheet theory of $N$ self-dual strings.
Here L/R signifies the fact the Kac-Moody symmetry is generated by the 
left/right 
chiral sector of the theory. 
We emphasis that the existence of this 
Kac-Moody symmetry is robust and is independent of supersymmetry or 
the details of the other part of the complete theory of 
the  self-dual strings. 

The existence of a Kac-Moody symmetry is intriguing. 
In the familiar case of the heterotic string, the existence of a group $\cG$ 
Kac-Moody symmetry in 
the left sector allows one to construct vertex operators which creates 
a Yang-Mills gauge symmetry $\cG$ in the 
spacetime. Now the spacetime of the self-dual strings is the
worldvolume of the M5-branes. Although
we don't have a vertex operator, it is tempting to speculate that the Kac-Moody 
symmetry will similarly create a set of gauge bosons in the
spacetime.  However since 
we do not have the vertex operators, it is not clear whether  a single
left (or right) handed Kac-Moody current is enough to create a spacetime 
gauge bosons, or whether the left and right handed  Kac-Moody currents must 
be taken together to create the gauge bosons.  This corresponds to having  
a gauge symmetry of $\cG\times \cG$ or $\cG$ on the system of 
$N$ coincident M5-branes. As we will see in section 2.2, the gauge
symmetry structure (equation \eq{GGA}) that is needed for the construction of
the non-abelian tensor gauge symmetry is different from the standard
direct-product structure and suggests that the mechanism for creating
the gauge bosons from the Kac-Moody current is different from the
standard case. 
In any case, the correspondence between worldsheet global symmetry and  
spacetime gauge symmetry should be a rather general statement.
All in all, we are motivated to 
conjecture that 
the (2,0) theory of a system of $N$ coincident M5-branes
is described by  a $U(N)\times U(N)$ tensor gauge symmetry algebra.

Below we will give an 
explicit construction for a theory of non-abelian tensor gauge fields 
based on a kind of   $G \times G$  gauge symmetry. The construction only
works with this $G\times G$ gauge symmetry structure and this is in support 
of the our conjecture that a $G\times G$ gauge symmetry is relevant for the 
description of multiple M5-branes.

\subsection{$G\times G$ tensor gauge symmetry for non-abelian 
tensor gauge field}
 
\nd\underline{Gauge and tensor gauge transformations}

Consider a gauge group $G \times G'$ where $G' =G$. Here 
$G$ is general and does 
not need to be $U(N)$. For notational convenience
we denote the second gauge group and the associated quantities with a prime. 
The spacetime 
dimension $D$ does not needed to be restricted to six.

Let $T^a$ be the generators of the Lie algebra $\frg$ of $G$, $a =1, 2,
\cdots, {\rm dim} \frg$. 
In addition to the  gauge fields $A^a_\mu, A'^a_\mu$,
we will include a 2-form tensor gauge field
$B_{\m\n}^a$ in the adjoint representation of $G$. Coupling to scalar fields
and fermions is easy and  will be considered in the 
next section. For now, we will concentrate on these fields.

Let us start with specifying the gauge transformations. 
We will take the gauge fields to  transform under $G\times G'$ as
\be \label{GGA}
\begin{array}{lll}
G: & \qquad \d_{\L} A_\mu = \del_\m \L+ [A_\m, \L], & 
\qquad\d_{\L} A'_\m= [A'_\m,\L],\\
G':& \qquad \d'_{\L'} A_\mu =  [A_\m, \L'], & \qquad
\d'_{\L'} A'_\m= \del_\m \L'+[A'_\m,\L'],
\end{array}
\ee
where $A_\mu = A_\m^a T^a$, $\L =\L^a T^a$, 
$A'_\mu = A'_\m{}^a T^a$, $\L' =\L'{}^a T^a$ are the 
Lie-algebra valued gauge fields 
and gauge parameters. Note that we have taken 
the gauge field $A$ (resp. $A'$) of the gauge group $G$ (resp. $G'$) 
to transform non-trivially in the adjoint representation 
of the other gauge group
$G'$ (resp. $G$). This is different from what one usually has in a
standard Yang-Mills theory. As we will explain below, 
that this is a consistent 
choice is entirely due to the presence of a tensor gauge symmetry 
in the theory. 

For the 2-form gauge fields, we will
take their gauge transformation  as 
\bea \label{B-g}
\d_{\L} B_{\m\n} = [B_{\m\n}, \L] + \frac{1}{2} \left(
[A'_\m , \del_\n \L] - [A'_\n, \del_\m \L]
\right), \nn \\
\d'_{\L'} B_{\m\n} = [B_{\m\n}, \L'] - \frac{1}{2} \left(
[A_\m , \del_\n \L'] - [A_\n, \del_\m \L']
\right).
\eea
It is convenient to introduce the field
\be
\cB_{\m\n} : = B_{\m\n} - \frac{1}{2}(F_{\m\n} -F'_{\m\n}),
\ee
where $F= dA + A^2$ and $F' =dA'+ A'{}^2$ are the ordinary gauge field strengths.
The field $\cB$ transforms covariantly under  $G \times G'$:
\bea
\d_{\L} \cB_{\m\n} = [\cB_{\m\n}, \L], \\
\d'_{\L'} \cB_{\m\n} = [\cB_{\m\n}, \L'].
\eea

In addition to the Yang-Mill gauge symmetry, there should also be a 
tensor gauge 
symmetry. In the case of a single tensor field, the tensor gauge
transformation takes the form
\be\label{t-abelian}
\d_{\L_\a} B_{\m\n} = \del_\m \L_\n -\del_\n \L_\m.  
\ee
The question is how this should be generalized for the non-abelian theory. We 
propose  the following tensor gauge transformations
\bea
\d_{\L_\a} B_{\m\n} &=& \frac{1}{2} \left[
(D_\m + D'_\m) \L_\n - (D_\n + D'_\n) \L_\m 
\right] \nn \\
&=& \left[
\del_\m + \frac{1}{2}(A_\m + A'_\m), \L_\n 
\right] - (\m \leftrightarrow \n),  \label{t1} \\
\d_{\L_\a} A_\m &=& \L_\m, \label{t2}\\
\d_{\L_\a} A'_\m &=& - \L_\m. \label{t3}
\eea
This implies that $\cB$ is tensor gauge invariant: 
\be
\d_{\L_\a} \cB_{\m\n} =0. 
\ee
In the free field limit where the commutator terms vanishes, the tensor gauge
transformation \eq{t1} decouples from the gauge fields $A, A'$ and 
reduces back to the \eq{t-abelian}.

The transformation properties of the field $\cB_{\m\n}$ makes itself a
convenient ingredient
for the construction of the 
covariant field strength. 
The field strength can be defined as
\be \label{cH}
\cH_{\m\n\l} \equiv [\cD_\m\; , \cB_{\n\l}] 
+ \mbox{($\m\n\l$ cyclic)}.
\ee
where  $\cD_\m = \del_\m+ [\cA_\m, \; \cdot]$
and $\cA_\m := A_\m+A'_\m$.
$\cH$ 
has the transformation properties
\bea
\d_\L \cH_{\m\n\l} &=& [\cH_{\m\n\l}, \L], \label{Ht1}\\
\d'_{\L'} \cH_{\m\n\l} &=& [\cH_{\m\n\l}, \L'], \\
\d_{\L_\a} \cH_{\m\n\l} &=& 0 \label{Ht3}
\eea
and satisfies the modified Bianchi identity
\be \label{bian}
\cD_{[\m} \cH_{\n\l\r]} = \frac{3}{2} [\cF_{[\m\n}, \cB_{\l\r]}],
\ee
where $\cF := d\cA + \cA^2$.

The above defined gauge transformations 
and tensor gauge transformations are consistent as they form a closed algebra. 
In fact, by acting on $B_{\m\n}, A_\m$ or $A'_\m$, it is easy to derive the 
following algebra of gauge and tensor gauge transformations:
\be
[\d_{\L^{(1)}_\m}, \d_{\L^{(2)}_\n}] =0, \label{gg0}
\ee
\be 
[\d_{\L_1}, \d_{\L_2} ] = \d_{[\L_1,\L_2]}, \label{gg1}
\ee
\be
[ \d'_{\L'_1}, \d'_{\L'_2} ] = \d'_{[\L'_1,\L'_2]},\label{gg2}
\ee
\be
[ \d_{\L}, \d'_{\L'} ] 
= \d_{\L_\m}+ \d_{\tilde{\L}} -\d'_{\tilde{\L}'} , \label{gg3}
\ee
\be
[\d_\L,\d_{\L_\a}] = - \d_{\tilde{\L}_\a},\label{gg4}
\ee
\be
[\d'_{\L'},\d_{\L_\a}] = - \d'_{\tilde{\L}'_\a}.\label{gg5}
\ee
Here in \eq{gg3},  
the parameter for the tensor gauge transformation on the right hand side 
is given by
\be
\L_\m \equiv \frac{1}{2}\left( [\del_\m \L, \L'] -[\L, \del_\m \L'] \right),
\ee
and the parameters $\tilde{\L}, \tilde{\L}'$ for the $G$ or $G'$ 
gauge transformations  on the right hand side  are
\be
\tilde{\L} \equiv \frac{1}{2} [\L, \L'], \quad 
\tilde{\L}' \equiv \frac{1}{2} [\L', \L];
\ee
while in \eq{gg4} and \eq{gg5}, the parameter for the 
tensor gauge transformation
on the right hand side are given by
\be
\tilde{\L}_\a \equiv [\L_\a, \L], \quad \tilde{\L}'_\a \equiv [\L_\a, \L'].
\ee

We note from \eq{gg3} 
that the commutator of a $G$-transformation and a $G'$-transformation 
results in a  tensor gauge transformation. This 
explains why our
proposed $G\times G$ gauge transformations \eq{GGA} has not been 
considered before in an ordinary 
Yang-Mills gauge theory since a tensor gauge symmetry is absent. 
We will refer to the algebra \eq{gg1}-\eq{gg5} of the gauge and tensor gauge 
transformations as the $G \times G$ tensor gauge symmetry structure 
for our non-abelian tensor gauge theory. 
 
For completeness, we remark that the 
above gauge 
symmetry algebra can also be written using a different basis in 
terms of  
the diagonal gauge field and the anti-diagonal gauge field
\be
\cA_\m := A_\m+ A'_\m, \quad \cC_\m := A_\m - A_\m'
\ee
and the diagonal and anti-diagonal gauge transformations
\be
\d^{(\rm d)}_\L:=\d_{\L/2} + \d'_{\L/2}, \quad
\d^{(\rm ad)}_\L:=\d_{\L/2} - \d'_{\L/2}. 
\ee
In terms of the these, the gauge transformation rules read
\be \label{CA}
\begin{array}{ccc}
\d^{(\rm d)}_\L \cA_\m = \del_\m \L + [\cA_\m, \L], &\qquad 
\d^{(\rm ad)}_\L\cA_\m =0, &\qquad 
\d_{\L_\a} \cA_\m =0, \\
\d^{(\rm d)}_\L \cC_\m = [\cC_\m, \L],  &\qquad 
\d^{(\rm ad)}_\L \cC_\m = \del_\m \L,  &\qquad 
\d_{\L_\a} \cC_\m = 2 \L_\m,
\end{array}
\ee
and
\bea
\d^{(\rm d)}_\L B_{\m\n} &=& [B_{\m\n}, \L]
- \frac{1}{4} \left(
[\cC_\m, \del_\n \L] - [\cC_\n, \del_\m \L] 
\right), \\ 
\d^{(\rm ad)}_\L B_{\m\n} &=& \frac{1}{4} \left(
[\cA_\m, \del_\n \L] - [\cA_\n, \del_\m \L] 
\right),\\
\d_{\L_\a} B_{\m\n} &=&  \left[
\del_\m + \frac{1}{2}\cA_\m, \L_\n 
\right] - (\m \leftrightarrow \n). \label{t1'}
\eea
Also it is
\be
\cB_{\m\n} = B_{\m\n} -\frac{1}{2}\left(
\big[\del_\m + \frac{1}{2} \cA_\m, \cC_\n\big] - (\m \leftrightarrow \n)
\right).
\ee

\nd\underline{Nature of the fields $\cC_\m$ and $\cA_\m$}

So far our construction involves the fields $\cA_\m$ and $\cC_\m$ in addition to 
the 2-form potential $B_{\m\n}$. The gauge field $\cC_\m$  is 
transformed by a shift under the tensor gauge 
transformation \eq{CA} and so can be gauged away if one fixes the tensor gauge
symmetry completely. Since part of the $B$-field can also be gauged away by
using the tensor gauge symmetry, that means if one does not want to  
introduce extra
pure gauge modes, 
one  should identify the field $\cC_\m$ with part of
$B_{\m\n}$. 
This can be achieved in a gauge invariant  way (with 
the   gauge symmetries as well as the tensor gauge symmetry all 
intact) by
imposing the constraint
\be \label{cond-B}
\cB_{\m 5} =0, \quad \m \neq 5,
\ee 
where 5 is an arbitrary fixed spacelike direction 
\footnote{One could equally take a timelike direction.} of 
the $D$-dimensional spacetime. Superficially the constraint \eq{cond-B} breaks
the Lorentz symmetry to $SO(D-2,1)$. 
But it is possible that the theory processes 
an additional modified Lorentz symmetry 
mixing the $5-\mu$ directions even if the theory is formulated with manifest
$(D-1)$-dimensional Lorentz invariance.
For example, this is the case in the Perry-Schwarz construction 
\cite{PS,schw1}  
of the single M5-brane theory. 
A special feature of the PS construction is that it is based on a 
$5\times 5$ 
tensor gauge fields with $B_{\mu \nu}$, $\mu =0,1,2,3,4$. The components
$B_{\mu 5}$ is completely missing in the formulation. This may appear
"artificial" but is in fact extremely natural in the manifestly Lorentz
covariant formulation of Pasti-Sorokin-Tonin (PST)
\cite{pst0,pst1,pst2} 
where the field 
$B_{\mu \nu}$ is extended to $B_{MN}$, $M =0, 1,2,3,4,5$. 
In addition an auxiliary   
scalar field $a$ is introduced with new gauge symmetries that allow one to
choose the gauge $B_{\mu 5} = 0$ and $a = x5$. In this gauge, the
Perry-Schwarz action is recovered. 
Recently, a non-abelian generalization of the Perry-Schwarz action 
was constructed in \cite{chua}. In this construction, a Yang-Mills gauge
symmetry $G=U(N)$ is present. 
It is envisaged that the full $G \times G$ formalism would be needed
in the PST-like formulation of the theory and the condition \eq{cond-B}
would then be a gauge fixing condition.

As for the nature of $\cA_\m$, there are two possibilities.
The first possibility is for the gauge field 
$\cA$ to be propagating. In this case, for a standard kinetic term $\Tr \cH^2$,
one see that  
the 2-form $B$ field
interacts with each other via the interaction through $\cA$. This is 
similar to the  familiar situation that with a standard kinetic term
$\psib \Ds \psi$ for fermions, $\psi$ interacts with each other 
only via the gauge field.
A more non-trivial  possibility is for 
the gauge field $\cA$ to be auxiliary and be
determined in terms of the other fields of the theory.
In this case there is direct nonlinear self interaction among the
2-form gauge field $B$ even within a single $\cH$; and 
our construction \eq{B-g}, \eq{t1} 
and \eq{cH} 
is a non-trivial generalization of the usual
non-abelian gauge transformation $\d A = d \L + [A,\L]$ and the
Yang-Mills field strength $F =dA +A^2$ which would be impossible to 
write down if only the field $B$ was allowed to appear. 
An example of these kind of constraint is
\be
X^2 \cF_{\m\n} = \cH_{\m\n\l} \cD^\l X, 
\ee
where $X$ is a scalar.

\section{Dynamics}

In this section, we discuss dynamics of  the 2-form gauge field. We
will also  include couplings to matter fields 
such as scalar fields and fermions and construct actions that are
invariant under the gauge and tensor gauge transformations. 
We
consider the generic case where the
construction is valid 
for general dimensions and self-duality for $\cH_{\m\n\l}$ is not assumed.
In the discussion section, we comment on the construction of the
theory of multiple M5-branes using our formalism of $G\times G$
tensor gauge  symmetry algebra.


\subsection{Matter couplings} 

The covariance and invariance \eq{Ht1} - \eq{Ht3} 
of the field strength $\cH$  allow one to write down
an invariant kinetic term $ \Tr \cH_{\m\n\l}^2$ immediately. 
Next let us include the coupling to matter fields. 
Consider first fields that are neutral under tensor
gauge transformation.
A simple example is for a  field $f = (f^a)$ 
to transform covariantly under the gauge transformations as
\be
\d^{\rm (d)}_\L f = [f, \L], \qquad \qquad 
\d^{\rm (ad)}_\L f = 0
\ee
and is invariant under the tensor gauge transformation
\be
\d_{\L_\a} f =0.
\ee
For these kind of fields, it is easy to construct their covariant
derivatives
\be
\cD_\m f = \del_\m + [\cA_\m, f].
\ee 
Note that these kind of matter fields
does not interact with the tensor gauge field minimally; non-minimal
interaction is possible, see \eq{action} below. 

It is also possible to include matters that are charged under the tensor 
gauge transformation. For simplicity take $G=U(N)$ and 
consider a  $N\times N$ Hermitian matrix of scalar field $\vphi$
which transforms under a tensor gauge transformation as
\be
\vphi \to U \vphi,
\ee
where $U = U(\L_\m)$ is some function of the tensor gauge parameter $\L_\m$.
Let us assume that $U$ does not depend on the derivatives of $\L_\m$
and has the form
\be
U = e^{\a^\m \L_\m},
\ee
for some matrix function $\a_\m$. Infinitesimally
\be \label{t-phi}
\d_{\L_\m} \vphi = \a^\m \L_\m \,\vphi.
\ee
It is convenient to introduce the field
\be
\tilde{\vphi} = (1+\b^\m \cC_\m) \vphi := C \vphi,
\ee
where $\b^\m\in U(1)$ and is independent of $\cC_\m$. The neutral case
is
included with $\b=0$. 
$\tilde{\vphi}$ is invariant under tensor gauge transformation if
\be
\b^\m + C \a^\m =0,
\ee
or
\be \label{ab}
\a^\m = - C^{-1} \b^\m.
\ee
This is well defined generically and the transformation \eq{t-phi} is a field
dependent one. 
As for the gauge transformations, we take them to be
\be \label{g-phi}
\d^{\rm (d)}_\L \vphi = [\vphi, \L], \qquad \qquad 
\d^{\rm (ad)}_\L \vphi = \a^\m \del_\m \vphi.
\ee 
It is easy to check that the transformations \eq{t-phi}, \eq{g-phi}
obey the algebra of transformations \eq{gg0} - \eq{gg5}. 

As a result, for a scalar field which transforms as \eq{t-phi} under the 
tensor gauge transformation and \eq{g-phi} under gauge
transformations, the covariant derivative
$\cD_\m \tilde{\vphi}$ is either invariant or covariant:
\be
\d_{\L_\m} (\cD_\n \tilde{\vphi} ) =0, \qquad
\d^{\rm (d)}_\L (\cD_\m \tilde{\vphi} ) = [\cD_\m \tilde{\vphi},\L],\qquad
\d^{\rm (ad)}_\L (\cD_\m \tilde{\vphi}) =0. 
\ee
Therefore the action
\be \label{S-phi}
S_{\vphi} = \int \Tr (\cD_\m \tilde{\vphi} )^2
\ee 
is invariant under the $G\times G$ gauge symmetry algebra. Notice that the 
coupling of $\vphi$ to $\cC_\m$ is rather non-standard. This is because
the gauge field
$\cC_\m$ does not transform in the standard way under the $G\times G$
gauge transformation, therefore it is not surprising that its coupling to the
matter fields is non-standard. 
We also note 
that the coupling of the field $\vphi$ to the tensor gauge field 
$B$ can either go
through $\cA$ in case $\cA$ is propagating, or nontrivially through 
$\cA$ as a 
function of $\cB$ in case $\cA$ is auxiliary. 
The construction for 
fermions goes in the same way.

We remark that 
the construction of gauge invariant coupling in ordinary 
Yang-Mills theory can also be 
proceeded 
by introducing an invariant field. Consider for example a scalar field $\vphi$ 
which transforms as $\vphi \to U^{-1} \vphi U$ under gauge transformation
$A_\m \to U^{-1} A_\m U + U^{-1} \del_\m U$. Introduce
\be
\tilde{\vphi} := W \vphi W^{-1},
\ee
where $W$ is a Wilson line which transforms as 
$W \to W U$ under gauge transformation. $\tilde{\vphi}$ is gauge
invariant.
It is
\be
\del_\m \tilde{\vphi} = W (D_\m \vphi) W^{-1}
\ee
and so the gauge invariant Lagrangian 
\be
\Tr (\del_\m \tilde{\vphi})^2 = \Tr(D_\m \vphi)^2
\ee
is indeed the same as the standard one constructed using covariant derivatives.
Our construction above for $\cH_{\m\n\l}$ and $\tilde{\vphi}$
was inspired and guided by this observation 
since
the direct construction of a covariant derivative for the tensor 
gauge transformation 
is met with immediate difficulty.

\subsection{Generic action in arbitrary dimensions}

Invariant action can be constructed readily. For example, an action
that is quadratic in the matter field is:
\bea \label{action}
S = \int d^D x \; \Tr \big[
&& (\cD_\m \tilde{\vphi})^2 + \bar{\tilde{\psi}} \G^\m \cD_\m \tilde{\psi} 
+ \frac{1}{4g_1^2}\cF_{\m \n}^2 
+ \frac{1}{g_2^2} \cB_{\m\n}^2 
+ \frac{1}{g_3^2} \cH_{\m\n\l}^2  \nn\\
&&
+ g_4 \cH_{\m\n\l}  \bar{\tilde{\psi}}  \G^{\m\n\l} \tilde{\psi} + 
g_5 \cF_{\m\n}  \bar{\tilde{\psi}}\G^{\m\n} \tilde{\psi}
\big].
\eea 
The mass dimensions of the fields are: $[A] =[A'] = 1, [B]=2, [\vphi]
=D/2 -1, [\psi] =(D-1)/2$ and for the couplings: $[g_1] =[g_2] =
2-D/2, [g_3] = 3-D/2, [g_4] = -2-D, [g_5] = -1-D$. 
It is straightforward to introduce a multiplet
of scalars and fermions to account for internal symmetry. Also one may
adjust the field content, the couplings and to include additional
terms such as Yukawa couplings to construct
supersymmetric action. In this action, the gauge field $\cA_\m$ is
propagating.

\subsection{A dual formulation of 5d Yang-Mills}

In this subsection, we demonstrate how our framework of non-abelian tensor
gauge symmetry could be used 
to construct a dual description of the 5d Yang-Mills
gauge theory in
terms of a three form field strength $\cH_{\m\n\l}$. Consider the
action of the non-abelian two-form $B_{\m\n}$ and one-form gauge field
$\cA_\m$  with the Lagrange multiplier field $\l_{\m\n}$,
\be \label{S5d}
S = \int \tr \left[
\cHt_{\m\n}^2 + (\cHt_{\m\n} - \cF_{\m\n}) \l_{\m\n}
\right]
,
\ee
where 
\be
\cHt^{\m\n} := \frac{1}{6} \e^{\m\n\r\l\s} \cH_{\r\l\s}
\ee
is the Hodge dual of $\cH_{\m\n\l}$ of \eq{cH}. Integrating out
$\l_{\m\n}$, we get the constraint
\be \label{c5d}
\cF_{\m\n} = \cHt_{\m\n}.
\ee
If we replace $\cH_{\m\n\l}$ in the integrated action 
using the constraint, we get the Yang-Mills 
description 
\be \label{S5d1}
S= S(\cA)= \int \tr \cF_{\m\n}^2.
\ee 
If we instead solve the constraint \eq{c5d} for $\cA_\m=
\cA_\m(\cB)$, then we get the description 
\be \label{S5d2}
S= S(\cB) = \int \tr \cHt_{\m\n}^2.
\ee
This provides an equivalent formulation of the Yang-Mills action in
terms of a non-abelian 2-form $B_{\m\n}$. The action is formulated in
terms of the tensor gauge invariant variables $\cB_{\m\n}$. 
Both \eq{S5d1} and \eq{S5d2} carries the same Yang-Mills gauge
symmetry.

\section{Discussions}

In this paper we have constructed a $G \times G$ tensor gauge symmetry 
algebra which 
includes the  gauge and tensor gauge transformations 
all together in a  natural way. 
This $G\times G$ structure allows us to write down readily a theory  of
non-abelian tensor gauge field in any dimensions, 
with or without self-duality, and with
matters neutral or charged under the tensor gauge symmetry. 
 
We have also shown that
one can construct an action of the 2-form gauge potential $B$
where the Yang-Mills gauge dynamics is completely fixed in 
terms of $B$ and the other matter fields. In this way, we
obtain a  3-form field strength with  non-trivial
non-linear terms of the 2-form gauge potential packaged nicely in
terms of a 1-form gauge field. Attempts to write down this formula
directly in the beginning by 
using $B$ only would be impossible. 
These are the main results of the paper.



Our construction of the non-abelian tensor gauge symmetry relies on 
the use of one-form gauge fields with a non-trivial kind of ``cross'' 
gauge transformations. It should be possible to construct 
higher tensor gauge theory with non-abelian symmetry in a similar
manner. These may be relevant for the studies of higher spin fields 
and duality \cite{hs1,hs2,hs3,hs4,hs5,hs6}.


\subsection{Comments on  M5-branes: self-duality}

With the application of M5-branes in mind,
the most interesting case is six dimensions. 
The low energy worldvolume theory of multiple M5-branes is given by
an interacting 2-form tensor gauge field theory with (2,0)
supersymmetry. The 3-form field strength has to satisfy a self-duality
condition as required by (2,0) supersymmetry. A  sensible
self-duality condition is of the form
\be \label{sd}
\cH^{-}_{\m\n\l} = \Phi_{\m\n\l},
\ee
where
\be \label{sd-parts}
\cH_{\m\n\l}^{\pm} := \frac{1}{2}(
\cH_{\m\n\l} \pm \cH^*_{\m\n\l}).
\ee
Here $\cH^*_{\m\n\l} = \frac{1}{6} \e_{\m\n\l\a\b\g} \cH^{\a\b\g}$ is
the Hodge dual
and $\cH^+$ (resp.  $\cH^-$) is  the self-dual (resp. anti-self-dual)
part of $\cH$. $\Phi_{\m\n\l}$ is a quantity which is required to
be tensor gauge invariant and is covariant under gauge transformation
in order for \eq{sd} to make sense. 
In the theory of multiple M5-branes, the self-duality
condition \eq{sd} is a part of the supermultiplet of equations of
motion and  $\Phi$ may  need to be non-trivial. 

In addition to a self-dual
$\cH$, the (2,0) supermultiplet of M5-branes  also contains 5 scalars 
and 8 fermions on-shell. The self-duality of the tensor gauge field makes it 
difficult to write down an action with standard $SO(5,1)$ 
Lorentz symmetry. The problem for a single M5-brane
with the self-duality equation of motion
\be \label{sd1}
\hat{\cH}_{\m\n\l} := \cH_{\m\n\l}- \cH_{\m\n\l}^* =0
\ee 
is solved with a formulation where the Lorentz symmetry is realized in  
a non-standard manner \cite{PS,schw1,pst0,pst1,pst2}. 
Consider the action
\be \label{SB}
S_B= \int d^6 x  \left(-\frac{1}{6} \cH_{\m\n\l} \cH^{\m\n\l} + 
\frac{1}{2 u^2} (u \hat{\cH)}_{\m\n} (u \hat{\cH})^{\m\n} \right),
\ee
where 
\be
(u \hat{\cH)}_{\m\n} = u^\r \hat{\cH}_{\r \m\n}
\ee
and $u^\m$ is a fixed constant vector in $U(1)$. The action has
manifest $SO(1,4)$ or $SO(5)$ Lorentz symmetry depending on whether $u$ is
spacelike or timelike. 
%
It is straightforward to check
that the action is invariant under
the symmetry 
\be \label{B-symm}
\d B_{\m\n} = \frac{1}{2} u_{[\m} \vphi_{\n]}, \quad \del_{[\m} u_{\n]}=0.
\ee
The equation
of motion for $B_{\m\n}$ is
\be \label{eom-B}
\e^{\m\n\l\r\a\b} \del_\m \left(
\frac{1}{u^2} u_\n (\hat{\cH} u)_{\l\r}
\right) =0.
\ee
One can show that the most general solution of it is of the form
\be \label{cH-soln}
 (\hat{\cH} u)_{\l\r} = u^2 \del_{[\l} \vphi_{\r]}+
u_{[\l} \del_{\r]} \vphi_\a u^\a +
u^\a \del_\a \vphi_{[\l} u_{\r]},
\ee
for an arbitrary function $\vphi_\m$. 
Since the right handed side of
\eq{cH-soln} has precisely the same form as the transformation of 
$ (\hat{\cH} u)_{\l\r}$ under the transformation \eq{B-symm} and so
one can use \eq{B-symm} to gauge fix it to zero
\be 
 (\hat{\cH} u)_{\l\r} =0.
\ee 
This implies the whole $\cH_{\m\n\l}$ is zero and so the equation of
motion \eq{eom-B} is exactly the same as the self-duality condition
\eq{sd1} after using the gauge symmetry. 
It turns out that for the non-abelian case, 
a similar action can be constructed 
which gives the self-duality equation as the equation of motion \cite{chua}.
One may also write down constraint on $\cA_\mu$ that determines its
in terms of $\cB_{\m\n}$ and other fields of the theory and leaves no local
degree of freedom in it. 
In this way, the non-abelian field strength 
$\cH_{\m\n\l}$ contains non-trivial  non-linear self interaction of 
the 2-form gauge potential packaged nicely in 
terms of the 1-form gauge field.
We note that non-propagating gauge fields also play an essential role in  
the BLG or ABJM theory for multiple M2-branes. There  the
gauge fields are auxiliary due to their Chern-Simons kinetic term  
and they are essential for a supersymmetric
construction. Here we would like to propose  that they are essential 
in the
construction of the
non-abelian tensor gauge theory of M5-branes. 
A detailed
discussion of the action principle for the self-duality equation 
and of the auxiliary nature of the gauge field $\cA_\m$
will be the subject of \cite{chua}.

We remark that another interesting approach \cite{He1,He2} makes use of canonical
variables has the advantage of being local, Lorentz invariant and
polynomial in the fields.
It will be interesting to
see whether one can use similar ideas to construct an action for a
non-abelian 2-form potential with a self-duality condition.

\subsection{Comments on  M5-branes: supersymmetry}

Apart from the self-duality equation of motion, the inclusion of (2,0) 
supersymmetry is another important aspect of the theory of multiple M5-branes.
Let us also comment on the supersymmetry. 
In analogy with the situation of the ABJM theory where the 
full $\cN=8$ supersymmetry is  supposed to be
seen only nonperturbatively after including the monopole sector, 
it might be possible that only
a  fraction of the (2,0) supersymmetry, i.e. (1,0) supersymmetry,  
is visible 
and full supersymmetry can be seen only
nonperturbatively. 
Therefore let us consider the possibility of constructing  
the theory of multiple
M5-branes using (1,0) supersymmetry.

In (1,0) supersymmetry, we expect 
the following supermultiplets to be useful:
\be
\begin{array}{ll}
\mbox{Tensor multiplet:}& \qquad(B^+_{\m\n}, X, \chi),\\
\mbox{Hyper-multiplet:} & \qquad (\phi_i, \psi),\\
\mbox{Yang-Mills multiplet:} & \qquad (A_\m, \l).
\end{array}
\ee
Here $B_{\m\n}$ has a self-dual 3-form field strength (hence the
superscript +), $X$ and $\phi_i
(i =1, \cdots 4) $ are scalar fields and $\chi, \l, \psi$ are
fermions. 
It is understood that all the fields carry the same non-abelian indices
$a$. With respect to (1,0) supersymmetry, a (2,0) tensor
multiplet is simply the sum of a (1,0) tensor multiplet and a (1,0) 
hyper-multiplet. However in order to have a  non-abelian tensor gauge
symmetry, it is necessary to have two additional gauge fields in
our construction. Therefore it  seems natural to use the
following (1,0) multiplets:
\be \label{20-10}
\mbox{1 $\times$ Tensor + 1 $\times$ Hyper + 2 $\times$ Yang-Mills }
\ee 
in the construction of the  theory of multiple
M5-branes  in the (1,0)
language. The Yang-Mills multiplet should be auxiliary and 
governed by a  suitable constraint.
The construction of the (2,0) non-abelian M5 theory in this way seems 
feasible \cite{chub}.

Recently, it has been proposed that 
the theory of multiple M5-branes compactified on a circle 
is nonperturbatively given by the D4-branes SYM theory by including 
instantons \cite{dou, lam2}. 
The fact that our $G\times G$ tensor gauge symmetry 
could be gauge fixed to a $G$ gauge symmetry provides a possible
connection of the gauge symmetries of the two theories which
may be useful in understanding this proposal. 
Very recently, a way to include the KK modes in the D4-branes
theory and an action of the multiple M5-branes
in $C$-field in terms of a 1-form gauge potential was proposed in \cite{cg}. 
As a theory of 1-form, there is no tensor gauge symmetry and the
proposed action has only a  $G=U(N)$ gauge symmetry, which 
is precisely the same symmetry we would obtain here
if the tensor gauge symmetry is fixed. Due to the coincidence of
symmetries, it is natural to 
wonder if this theory admits a dual formulation in terms of our
framework $G \times G$ tensor gauge field.
It will be very interesting to clarify the possible connection.

\section*{Acknowledgements}

It is a pleasure to thank Paul Heslop, Pei-Ming Ho,   Costis
Papageorgakis, Gurdeep Sehmbi,
and in particular Douglas Smith for useful discussions and
comments on the manuscript. 

\end{document}